\documentclass{article}
\usepackage{spconf,amsmath,graphicx}
\usepackage{graphicx,wrapfig,lipsum}

\usepackage{amsmath}
\def\sup#1{$^{#1}$}
\usepackage{amssymb}

\usepackage{booktabs}
\usepackage{multirow}
\usepackage{multicol}

\usepackage{graphicx}
\usepackage{color}
\usepackage{float}
\restylefloat{table}
\usepackage{placeins}
\usepackage{epstopdf}

\usepackage{hyperref}
\DeclareMathOperator{\Tr}{Tr}
\usepackage{colortbl}
\usepackage{arydshln,graphicx,xcolor,array}
\usepackage{tikz}
\usetikzlibrary{matrix}

\usepackage{subfig}
\usepackage{fancyhdr}

\title{Target And Background Separation in Hyperspectral Imagery for Automatic Target Detection}
\name{Ahmad W. Bitar\sup{1}, Loong-Fah Cheong\sup{2} and Jean-Philippe Ovarlez\sup{1,3}}
\address{\sup{1}SONDRA/CentraleSup\'elec, Plateau du Moulon, 3 rue Joliot-Curie, F-91190 Gif-sur-Yvette, France\\
\sup{2}National University of Singapore (NUS), Singapore, Singapore\\
\sup{3}ONERA, DEMR/TSI, Chemin de la Huni\`ere, 91120 Palaiseau, France}

\fancypagestyle{mahmood}{%
   \fancyhf{} % clear all fields
   
   \fancyhead[C]{\color{red}This paper has been submitted to IEEE ICASSP'18 in October 2017 and \underline{got accepted for publication} in 29 January 2018. The paper has been presented at the conference in Calgary in 18 April 2018}
}%

\makeatletter
\let\ps@IEEEtitlepagestyle\ps@mahmood
\makeatother
\begin{document}
\ninept
\maketitle
\begin{abstract}
%In this paper, based on a modification of the Robust Principal Component Analysis (RPCA), we regard the given hyperspectral image (HSI) as being made up of the sum of low-rank background HSI and a sparse target HSI. By assuming that the target's spectra is known, we aim to cast the 

In this paper, we propose a method for separating known targets of interests from the background in hyperspectral imagery. More precisely, we regard the given hyperspectral image (HSI) as being made up of the sum of low-rank background HSI and a sparse target HSI that contains the known targets based on a pre-learned target dictionary specified by the user. Based on the proposed method, two strategies are outlined and evaluated independently to realize the target detection on both synthetic and real experiments. 
%Both synthetic as well as real experiments are performed to gauge the target detection performances of the two strategies.   

\end{abstract}
\section{Introduction}
\label{sec:intro}

A hyperspectral image (HSI) is a three dimensional data cube consisting of a series of images of the same spatial scene in a contiguous and multiple narrow spectral wavelength (color) bands \cite{Shaw02, manolakis2003hyperspectral, manolakis_lockwood_cooley_2016}. Each pixel in the HSI is a $p$-dimensional vector, $\mathbf{x} \in \mathbb{R}^p$, where $p$ stands for the total number of spectral bands. With the rich information afforded by the high spectral dimensionality, target detection is not surprisingly one of the most important applications in hyperspectral imagery \cite{Shaw02, manolakis2003hyperspectral, Manolakis14, Manolakis09, manolakis2002detection, Frontera13, Frontera14}. 
Usually, the detection is built using a binary hypothesis test that chooses between the following competing null and alternative hypothesis: target absent ($H_0$), that is, the test pixel $\mathbf{x}$ consists only of background; and target present ($H_1$) where $\mathbf{x}$ may be either fully or partially occupied by the target material. We can regard each test pixel $\mathbf{x}$ as being made up of $\mathbf{x} = \alpha \mathbf{t}$ + $(1-\alpha) \mathbf{b}$, where $0 \leq \alpha \leq 1$ is the target fill-fraction, $\mathbf{t}$ is the spectrum of the target, and $\mathbf{b}$ the spectrum of the background.

Different target detectors (e.g., Matched Filter \cite{Manolakis00, Nasrabadi08}, Normalized Matched Filter \cite{kraut1999cfar}, Kelly detector \cite{4104190}) have been developed and which are dependent on the target spectra. These classical target detectors present several limitations.
Firstly, the dependency on the unknown covariance matrix (of the background surrounding the test pixel) whose entries have to be carefully estimated specially in large dimensions \cite{LEDOIT2004365, AhmadCamsap2017} and to ensure success under different environment \cite{5606730, 6884641, 6894189}.
Secondly, there is always an explicit assumption on the statistical distribution characteristics of the observed data.
Lastly, the use of only a single reference spectrum for the target of interest may be inadequate since in real world hyperspectral imagery, various effects that produce variability to the material spectra (e.g., atmospheric conditions, sensor noise, material composition, etc.) are inevitable. For instance, target signatures are typically measured in laboratories or in the field with hand-held spectrometers that are at most a few inches from the target surface. Hyperspectral images, however, are collected at huge distances away from the target and have significant atmospheric effects present.

To more effectively separate these non-Gaussian noise from signal, and to have a target detector that is invariant to atmospheric effects, dictionaries of target and background have been developed (denoted as $\mathbf{A}_t$ and $\mathbf{A}_b$ in this paper) and the test signal is then modeled as a sparse linear combination of the prototype signals taken from the dictionaries \cite{chen11, Chen11b, Zhang15}. This sparse representation approach can alleviate the spectral variability caused by atmospheric effects, and can also better deal with a greater range of noise phenomena. Our work falls under this broad family of dictionary-based approach.
Although these dictionary-based-methods can in principle address all the aforementioned limitations, the main drawback is that they usually lack a sufficiently universal dictionary, especially for the background $\mathbf{A}_b$; some form of in-scene adaptation would be desirable. Chen {\it et al.} \cite{chen11, Chen11b} have demonstrated in their sparse representation approach, that using an adaptive scheme (a local method) to construct $\mathbf{A}_b$ usually yields better target detection results than with a global dictionary generally constructed from some background materials (e.g., trees, grass, road, buildings, vegetation, etc.). This is to be expected since the subspace spanned by the background dictionary $\mathbf{A}_b$ becomes adaptive to the local statistics.
Zhang {\it et al.} \cite{Zhang15} were based on the same adaptive scheme in their sparse representation-based binary hypothesis (SRBBH) approach. 

In \cite{chen11, Chen11b, Zhang15}, the adaptive scheme is based on a dual concentric window centered on the test pixel (see Figure {\bf1(a)}), with an inner window region (IWR) centered within an outer window region (OWR), and only the pixels in the OWR will constitute the samples for $\mathbf{A}_b$. Clearly, the dimension of IWR is very important and has a strong impact on the target detection performance since it aims to enclose the targets of interests to be detected. It should be set larger than or equal to the size of all the desired targets of interests in the corresponding HSI, so as to exclude the target pixels from erroneously appearing in $\mathbf{A}_b$. However, information about the target size in the image is usually not at our disposal. It is also very unwieldy to set this size parameter when the target could be of irregular shape (e.g., searching for lost plane parts of a missing aircraft). Another tricky situation is when there are multiple targets in close proximity in the image (e.g., military vehicles in long convoy formation). 

%\begin{figure}[!tbp]
%\centering
%\begin{minipage}[b]{0.245\textwidth}
%\includegraphics[width=\textwidth]{dual_window2.png}
%\caption{Dual concentric window}
%\label{fig:dual_window}
%\end{minipage}
%\end{figure}

In this paper, we address all the aforementioned challenges in constructing $\mathbf{A}_b$ by providing a method capable of automatically removing the targets from the background, and hence, avoiding the use of an IWR to construct $\mathbf{A}_b$ as well as dealing with a larger range of target size, shape, number, and placement in the image. Based on a modification of the recently developed Robust Principal Component Analysis (RPCA) \cite{Candes11}, our method decomposes an input HSI into a background HSI (denoted by $\mathbf{L}$) and a sparse target HSI (denoted by $\mathbf{E}$) that contains the targets with the background is suppressed.
\\
While we do not need to make assumptions about the size, shape, or number of the targets, our method is subject to certain generic constraints that make less specific assumption on the background or target. These constraints are similar to those used in RPCA \cite{Candes11, NIPS2009_3704}, including: 1) the background is not too heavily cluttered with many different materials with multiple spectra, so that the background signals should span a low-dimensional subspace, a property that can be expressed as the low rank condition of a suitably formulated matrix \cite{ChenYu, Zhang15b, Ahmad2017a}; 
2) the total image area of all the target(s) should be small relative to the whole image (i.e. spatially sparse), though there is no restriction on target shape or the proximity between targets.

Our method further assumes that the target spectra is available to the user and that the atmospheric influence can be accounted for by the target dictionary $\mathbf{A}_t$. This pre-learned dictionary $\mathbf{A}_t$ is used to cast the general RPCA into a more specific form, specifically, we further factorize the sparse component $\mathbf{E}$ from RPCA into the product of $\mathbf{A}_t$ and a sparse activation matrix $\mathbf{C}$. This modification is essential to disambiguate the true targets from other small objects, as the following discussion will show. In our application, there are often other small, heterogeneous, high contrast regions that are non-targets. These would have been deemed as outliers (targets) under the general RPCA framework. Compounding the decomposition is also the often uniform material present in most targets, which means that they would contribute only a small increase in the rank of the background HSI if they were to be grouped in the background HSI. Indeed, some other heterogeneous non-target objects or specular highlights may contribute a larger increase in rank and thus they are more liable to be treated as outliers under general RPCA. 

\begin{figure}[!tbp]
\minipage{0.12\textwidth}
\includegraphics[width=\textwidth]{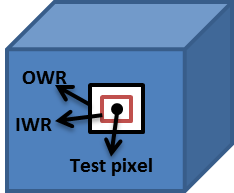}
{\begin{center}%
\vspace{-3mm}
{\bf (a)}
\end{center}}
\endminipage\hfill
\minipage{0.121\textwidth}
\includegraphics[width=\linewidth]{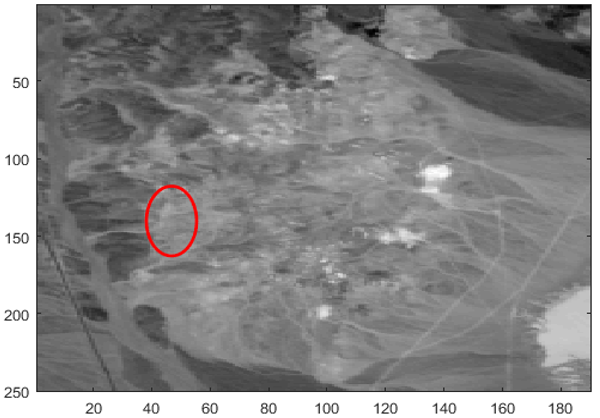}
{\begin{center}%
\vspace{-3mm}
{\bf (b)}
\end{center}}
\endminipage\hfill
\minipage{0.121\textwidth}
\includegraphics[width=\linewidth]{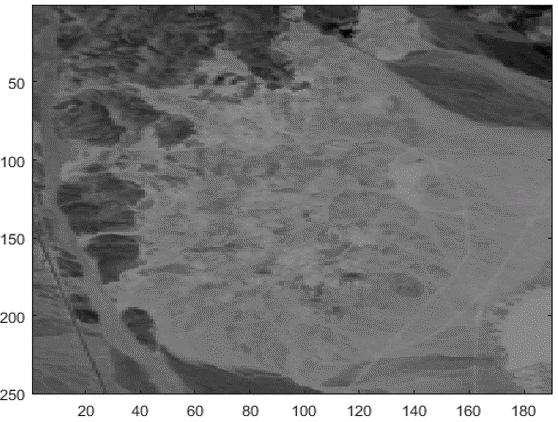}
{\begin{center}%
\vspace{-3mm}
{\bf (c)}
\end{center}}
\endminipage\hfill
\minipage{0.121\textwidth}
\includegraphics[width=\linewidth]{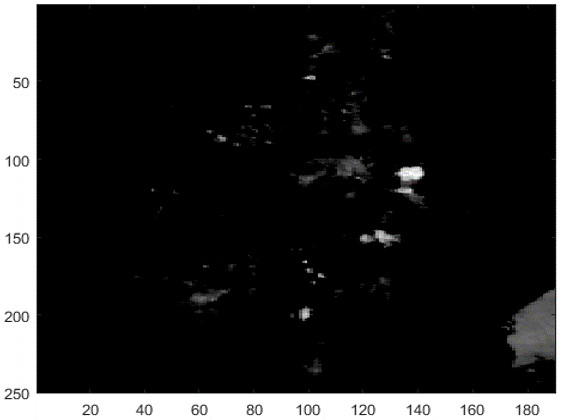}
{\begin{center}%
\vspace{-3mm}
{\bf (d)}
\end{center}}
\endminipage
\vspace{-3mm}
\caption{{\bf(a)}: The dual concentric window method. {\bf(b)}: Original Cuprite HSI (the buddingtonite target pixels are located inside the red ellipse). {\bf(c)}: Low rank background HSI $\mathbf{L}$. {\bf(d)}: Sparse target HSI $\mathbf{E}$ (after some thresholding).} %Note that in {\bf 1(b) 1(c) 1(d)}, we exhibit the mean power in dB over the 186 spectral bands.}
\label{fig:SPCP_example}
\end{figure}

Let us take an example in Figure \ref{fig:SPCP_example} (see Figure {\bf1(b) 1(c) 1(d)}) that uses the RPCA model solved via Stable Principal Component Pursuit \cite{SPCPCandes} on a region of the Cuprite mining district area \cite{Swayze10245, JGRE:JGRE1642} in which are present some Buddingtonite pixels considered as targets. As can be seen, despite the effort to individually tune the parameters for best separation, it is not possible to obtain a clean separation. The RPCA model was not able to find the Buddingtonite target pixels but instead other small heterogeneous and high contrast regions are preferentially deposited in the sparse target image $\mathbf{E}$.

In this regard, the incorporation of the target dictionary prior can, we feel, greatly help in identifying the true targets and separate them from the background. From the proposed method, we use the background HSI $\mathbf{L}$ for a more accurate construction of $\mathbf{A}_b$, following which various dictionary-based-methods can be used to carry out a more elaborate binary hypothesis test. Via the background HSI $\mathbf{L}$, a locally adaptive $\mathbf{A}_b$ can be constructed without the need of using an IWR, and also avoiding contamination by the target pixels.
\\
An alternative strategy would be to directly use the target HSI (the product of $\mathbf{A}_t$ and the sparse activation matrix $\mathbf{C}$) as a detector. That is, we detect the non-zero entries of the sparse target image, and targets are deemed to be present at these non-zero support. 

{\em Main Notations:} The notation $(.)^T$ and $\mathrm{Tr}$ stand for the transpose and trace of a matrix, respectively. In addition, $\mathrm{rank}(.)$ is for the rank of a matrix. A variety of norms on matrices will be used. For instance, $\mathbf{M}$ is a matrix, $[\mathbf{M}]_{:,\,j}$ is the $j$-th column. The matrix $l_{2,0}$, $l_{2,1}$ norms are defined by $||\mathbf{M}||_{2,0} = \# \{ j \, : \, ||[\mathbf{M}]_{:,\,j}||_2 \, \not= \, 0\}$, $||\mathbf{M}||_{2,1} = \sum_j ||[\mathbf{M}]_{:,\,j}||_2$, respectively. The Frobenius norm and the nuclear norm (the sum of singular values of a matrix) are denoted by $||\mathbf{M}||_F$ and $||\mathbf{M}||_* = \Tr(\sqrt{\mathbf{M}^T \mathbf{M}})$, respectively.
%\footnote{More real examples demonstrating the inadequacy of RPCA is found in our ArXiv paper \cite{Ahmad2017a}.}

\vspace{-2mm}
\section{Main contributions}
\label{sec:contributions}

\subsection{Problem formulation} 
Suppose a HSI of size $h \times w \times p$, where $h$ and $w$ are the height and width of the image scene, respectively, and $p$ is the number of spectral bands. 
Consider that the given HSI contains $q$ pixels $\{ \mathbf{x_i}\}_{i\in[1,\,q]}$ of the form: $\mathbf{x}_i = \alpha_i \mathbf{t}_i + (1 - \alpha_i)\mathbf{b}_i$ with $0<\alpha_i \leq 1$, where $\mathbf{t}_i$ represents the known target that replaces a fraction $\alpha_i$ of the background $\mathbf{b}_i$ (i.e. at the same spatial location). The remaining ($e-q$) pixels in the given HSI, with $e = h \times w$, are thus only background ($\alpha=0$).
By assuming that all $\{\mathbf{t}_i\}_{i\in[1,\,q]}$ consist of similar materials, thus they should be represented by a linear combination of $N_t$ common target samples $\{\mathbf{a}^t_{j}\}_{j \in [1, \, N_t]}$, where $\mathbf{a}_j^t \in\mathbb{R}^p$ (the superscript $t$ is for target), but weighted with different set of coefficients $\{\beta_{i,j}\}_{j\in[1, N_t]}$. Thus, each of the $q$ pixels is represented as:
\begin{align}
\footnotesize
\mathbf{x}_i = \alpha_i \sum\limits_{j=1}^{N_t} \Big(\beta_{i,j}\mathbf{a}^t_{j} \Big) + (1 - \alpha_i) \mathbf{b}_i \hspace{0.5cm} i \in [1,q]\, . 
\end{align}
We rearrange the given HSI into a two-dimensional matrix $\mathbf{D}\in \mathbb{R}^{e \times p}$, with $e = h \times w$ (by lexicographically ordering the columns). This matrix $\mathbf{D}$, can be decomposed into a low rank matrix $\mathbf{L}_0$ representing the pure background, a sparse matrix capturing any spatially small signals residing in the known target subspace, and a noise matrix $\mathbf{N}_0$. More precisely, the model used is 
\begin{align}\label{eq:mod2}
\footnotesize
\mathbf{D} &= \mathbf{L}_0 + (\mathbf{A}_t\mathbf{C}_0)^T + \mathbf{N}_0\,,
\end{align}
where $(\mathbf{A}_t\mathbf{C}_0)^T$ is the sparse target matrix, ideally with $q$ non-zero rows representing $\alpha_i\mathbf{t}^T_i, ~{i\in[1,q]}$ , with target dictionary $\mathbf{A}_t \in \mathbb{R}^{p \times N_t}$ having columns representing target samples $\{\mathbf{a}^t_{j}\}_{j \in [1, N_t]}$, and coefficient matrix $\mathbf{C}_0\in\mathbb{R}^{N_t \times e}$ that should be a sparse column matrix, again ideally containing $q$ non-zero columns each representing $\alpha_i [\beta_{i,1}, \, \cdots, \, \beta_{i, N_t}]^T$, $i \in [1, q]$. $\mathbf{N}_0$ is assumed to be independent and identically distributed Gaussian noise with zero mean and unknown standard deviation.
\\
After reshaping $\mathbf{L}_0$, $(\mathbf{A}_t\mathbf{C}_0)^T$ and $\mathbf{N}_0$ back to a cube of size $h \times w \times p$, we call these entities the ``low rank background HSI'', ``sparse target HSI'', and ``noise HSI'', respectively.
\\
In order to recover the low rank matrix $\mathbf{L}_0$ and sparse target matrix $(\mathbf{A}_t\mathbf{C}_0)^T$, we consider the following minimization problem:
\vspace{-0.8mm}
\begin{align}\label{eq:our_minimization}
\footnotesize
\underset{\mathbf{L}, \mathbf{C}} {\mathrm{min}} \,  \Bigl\{\tau \, \mathrm{rank}(\mathbf{L})+ \lambda \, ||\mathbf{C}||_{2,0} +  ||\mathbf{D} - \mathbf{L} - (\mathbf{A}_t\mathbf{C})^T||_F^2 \Bigr\}\,,
\end{align}
where $\tau$ controls the rank of $\mathbf{L}$, and $\lambda$ the sparsity level in $\mathbf{C}$.

\subsection{Solving our problem by convex optimization}
Problem \eqref{eq:our_minimization} is NP-hard due to the presence of the rank term and the $||.||_{2,0}$ term. We relax these terms to their convex proxies, specifically, using nuclear norm $||\mathbf{L}||_*$  as a surrogate for the rank$(\mathbf{L})$ term, and the $l_{2,1}$ norm for the $l_{2,0}$ norm.
We now need to solve the following convex minimization problem:
\vspace{-0.8mm}
\begin{align}{\label{eq:convex_model}}
\footnotesize
\underset{\mathbf{L}, \mathbf{C}} {\mathrm{min}} \, \Bigl\{\tau \,||\mathbf{L}||_*+ \lambda \,||\mathbf{C}||_{2,1} +  ||\mathbf{D} - \mathbf{L} - (\mathbf{A}_t \mathbf{C})^T||_F^2 \Bigr\}\,,
\end{align}
Problem \eqref{eq:convex_model} is solved via an alternating minimization of two subproblems. Specifically, at each iteration $k$:
\begin{subequations}\label{eq:sub}
\footnotesize
\begin{alignat}{2}
\label{eq:sub1}
\mathbf{L}^{(k)} &= \underset{\mathbf{L}} {\mathrm{argmin}} \, \Bigl\{||\mathbf{L} - \big(\mathbf{D} - (\mathbf{A}_t \mathbf{C}^{(k-1)})^T\big)||_F^2 + \tau \,||\mathbf{L}||_* \, \Bigr\}\,, \\
\label{eq:sub2}
\mathbf{C}^{(k)} &= \underset{\mathbf{C}} {\mathrm{argmin}} \, \Bigl\{||(\mathbf{D} - \mathbf{L}^{(k)})^T -  \mathbf{A}_t \mathbf{C}||_F^2 + \lambda \,||\mathbf{C}||_{2,1} \, \Bigr\}.
\end{alignat}
\end{subequations}
The minimization sub-problems \eqref{eq:sub1} \eqref{eq:sub2} are convex and each can be solved optimally. \eqref{eq:sub1} is solved via the Singular Value Thresholding operator \cite{SVT2010}. \eqref{eq:sub2} refers to the Lasso problem (if we reshape the matrix $\mathbf{C}$ into a vector) which can be solved by various methods, among which we adopt the Alternating Direction Method of Multipliers (ADMM) \cite{Boyd:2011:DOS:2185815.2185816}. 
More precisely, we introduce an auxiliary variable $\mathbf{F}$ into sub-problem \eqref{eq:sub2} and recast it into the following form:
\begin{equation}
\label{eq:problemm}
\footnotesize
%\begin{split}
(\mathbf{C}^{(k)}, \mathbf{F}^{(k)}) = \underset{s.t.~~ \mathbf{C} = \mathbf{F}} {\mathrm{argmin}} \, \Bigl\{||(\mathbf{D} - \mathbf{L}^{(k)})^T - \mathbf{A}_t \mathbf{C}||_F^2 + \lambda \, ||\mathbf{F}||_{2,1}\Bigr\}
%\\
%s.t.~~ \mathbf{C} = \mathbf{F}
%\end{split}
\end{equation}
%This problem can then be solved via the Augmented Lagrangian Multiplier method \cite{Lintheaugmented, HiSm1} as follows:
Problem \eqref{eq:problemm} is then solved as follows (scaled form of ADMM):
%\begin{subequations}
%\begin{align}
%\footnotesize
%\begin{split}
%\mathbf{C}^{(k)} = \underset{\mathbf{C}} {\mathrm{argmin}} \, \Bigl\{||(\mathbf{D} - \mathbf{L}^{(k)})^T - \mathbf{A}_t \mathbf{C}||_F^2
%\\
%+ \frac{\rho^{(k-1)}}{2}\,||(\mathbf{C} - \mathbf{F}^{(k-1)} - \frac{1}{\rho^{(k-1)}}\mathbf{Z}^{(k-1)}||_F^2 \, \Bigr\}
%\\
%\mathbf{F}^{(k)} = \underset{\mathbf{F}} {\mathrm{argmin}} \, \Bigl\{\lambda \, ||\mathbf{F}||_{2,1} + \frac{\rho^{(k-1)}}{2}\,||(\mathbf{C}^{k} - \mathbf{F} - \frac{1}{\rho^{(k-1)}}\mathbf{Z}^{(k-1)}||_F^2 \, \Bigr\}
%\\
%\mathbf{Z}^{(k)} = \mathbf{Z}^{(k-1)} + \rho^{(k-1)} \,(\mathbf{C}^{(k)} - \mathbf{F}^{(k)})
%\end{split}
%\end{align}
\begin{subequations}
\footnotesize
\begin{align}
\begin{split}
\mathbf{C}^{(k)} = \underset{\mathbf{C}} {\mathrm{argmin}} \, \Bigl\{||(\mathbf{D} - \mathbf{L}^{(k)})^T - \mathbf{A}_t \mathbf{C}||_F^2
\\
+ \frac{\rho^{(k-1)}}{2}\,||(\mathbf{C} - \mathbf{F}^{(k-1)} + \frac{1}{\rho^{(k-1)}}\mathbf{Z}^{(k-1)}||_F^2 \, \Bigr\}
\end{split}
\\
\mathbf{F}^{(k)} &= \underset{\mathbf{F}} {\mathrm{argmin}} \, \Bigl\{\lambda \, ||\mathbf{F}||_{2,1} + \frac{\rho^{(k-1)}}{2}\,||(\mathbf{C}^{k} - \mathbf{F} + \frac{1}{\rho^{(k-1)}}\mathbf{Z}^{(k-1)}||_F^2 \, \Bigr\}
\\
\mathbf{Z}^{(k)} &= \mathbf{Z}^{(k-1)} + \rho^{(k-1)} \,(\mathbf{C}^{(k)} - \mathbf{F}^{(k)})
\end{align}
\end{subequations}
where $\mathbf{Z} \in \mathbb{R}^{N_t \times e}$ is the Lagrangian multiplier matrix, and $\rho$ is a positive scalar. We initialize $\mathbf{L}^{(0)} = \mathbf{C}^{(0)} = \mathbf{F}^{(0)} = \mathbf{Z}^{(0)} = \boldsymbol{0}$, $\rho^{(0)} = 10^{-4}$ and update $\rho^{(k)} = 1.1 \, \rho^{(k-1)}$. The criteria for convergence of sub-problem \eqref{eq:sub2} is $||\mathbf{C}^{(k)} - \mathbf{F}^{(k)}||_F^2 \leq 10^{-6}$.
\\
For Problem \eqref{eq:convex_model}, we stop the iteration when the following convergence criterion is satisfied:
\\
\\
$
~~~~\frac{||\mathbf{L}^{(k)} - \mathbf{L}^{(k-1)}||_F}{||\mathbf{D}||_F} \leq \epsilon~~~~~\text{and} ~~~~~
\frac{||(\mathbf{A}_t \mathbf{C}^{(k)})^T - (\mathbf{A}_t \mathbf{C}^{(k-1)})^T||_F}{||\mathbf{D}||_F} \leq \epsilon
$
\\
\\
where $\epsilon>0$ is a precision tolerance parameter. We set $\epsilon = 10^{-4}$.

%\subsection{Target dictionary construction} 
%\label{sec:dict_const}
%An important problem that requires very careful attention is the construction of an appropriate dictionary $\mathbf{A}_t$ in order to well capture the target to separate from the background.
%In reality, the target present in the HSI can be highly affected by the atmospheric conditions, sensor noise, material composition, etc., that may produce huge variations on the target's spectra. In view of these real effects, it is very difficult to model the target dictionary ($\mathbf{A}_t$) well. But this raises the question on how these effects should be dealt with.
%Some scenarios for modelling the target dictionary have been followed over several decades. For example, by using physical models and the MORTRAN atmospheric modeling program \cite{Berk89}, target spectral signatures can be generated under various atmospheric conditions. For simplicity, in this paper, we handle this problem by exploiting target samples available in some online spectral libraries. More precisely, $\mathbf{A}_t$ can be constructed via the United States Geological Survey (USGS - Reston) Spectral Library \cite{Clark93}. However, the user can also deal with the Advanced Spaceborne Thermal Emission and Reflection (ASTER) spectral library \cite{Baldridge09} that includes data from the USGS Spectral Library, the Johns Hopkins University (JHU) Spectral Library, and the Jet Propulsion Laboratory (JPL) Spectral Library.   

\vspace{-1.5mm}

\subsection{What after the target and background separation} 
\label{sec:what_after}
\vspace{-1.5mm}
Two strategies are available to us to realize the target detection\footnote{More information about this work are in our ArXiv version \cite{Ahmad2017b}. Note that in the synthetic and real experiments later, both the HSI and the target samples are normalized to values between 0 and 1.}.

{\bf Strategy one}: We use the background HSI $\mathbf{L}$ for a more accurate construction of $\mathbf{A}_b$. For each test pixel in the original HSI, we create a concentric window of size $m \times m$ on the background HSI $\mathbf{L}$, and all the pixels within the window (except the center pixel) will each contribute to one column in $\mathbf{A}_b$. Note that this concentric window amounts to an OWR of size $m \times m$ with IWR of size $1 \times 1$. Next, we make use of the SRBBH detector \cite{Zhang15}, but with the background dictionary $\mathbf{A}_b$ constructed in the preceding manner. Note that for this scheme to work, we do not need a clean separation (by clean separation, we mean that all targets are present in $(\mathbf{A}_t\mathbf{C})^T$ with no false alarms); specifically, we require the entire target fraction to be separated from the background and deposited in the target image, but some of the background objects can also be deposited in the target image. As long as enough signatures of these background objects remain in the background HSI $\mathbf{L}$, the $\mathbf{A}_b$ constructed will be adequately representative of the background. 

{\bf Strategy two}: We use $(\mathbf{A}_t \mathbf{C})^T$ directly as a detector. Note that for this scheme to work, we require as few false alarms as possible to be deposited in the target image, but we do not need the target fraction to be entirely removed from the background (that is, a very weak target separation can suffice). As long as enough of the target fractions are moved to the target image such that non-zero support is detected at the corresponding pixel location, it will be adequate for our detection scheme. From this standpoint, we should choose a $\lambda$ that is relatively large, so that the target image is really sparse with zero or little false alarms, and only signals that reside in the target subspace specified by $\mathbf{A}_t$ will be deposited there.

%\footnote{The reason why we focus on the SRBBH detector instead of \cite{chen11} is because it combines the idea of binary hypothesis and sparse representation, obtaining a more complete and realistic model than \cite{chen11}.}

\vspace{-1.5mm}

\section{Experiments and analysis}
\label{sec:global_synthetic_Experiments}
\vspace{-1.5mm}
In this section, we perform both synthetic as well as real experiments to gauge the target detection performances of the two preceding strategies in Subsection \ref{sec:what_after}. The evaluations are done on two small zones acquired from the online Cuprite HSI data \cite{CupriteHSIOnline}. The Cuprite HSI  is a mining district area \cite{Swayze10245, JGRE:JGRE1642} containing well exposed zones of advanced argillic alteration, consisting principally of kiolinite, alunite, and hydrothermal silica. It consists of 224 spectral (color) bands. Prior to some analysis of the Cuprite HSI, the spectral bands 1-4, 104-113 and 148-167 are removed due to the water absorption in those bands. As a result, a total of 186 bands are used.

{\bf \em Parameters settings}:
For Strategy one, we found that the ratios of $\tau$ to $\lambda$ should be high to make sure that all of the targets are removed to the target image. We set this ratio to approximately $6$ for the synthetic experiments and $10$ for the real experiments. The ratio for the latter case must be higher because for the real experiments, we do not really have a comprehensive enough target dictionary to represent the target well and thus we need extra incentive for the target fractions to go to the target image. Now for Strategy two, we found that the ratio of $\tau$ to $\lambda$ must be equal to $\frac{5}{2}$ in both synthetic and real experiments.
In fact, the different requirements imposed by the two strategies that lead to our particular choice of the $\tau$ to $\lambda$ ratio also dictate how we should set the relative values of the weights between the first two terms and the third term in \eqref{eq:convex_model}. A lower penalty associated with the third term (that is, by raising the absolute levels of $\tau$ and $\lambda$) would tolerate more deviation and thus encourage more noise or image clutters (by image clutters we mean the small heterogeneous objects and specular highlights) to be absorbed by this term. This is particularly important for Strategy two when there are a lot of image clutters that do not exactly conform to a low rank background model: since these clutters do not satisfying the low rank property, they have a propensity to show up in the second term if we do not sufficiently lower the penalty for the third term, and thus, contribute to a lot of false alarms for Strategy two. On the other hand, such a low-penalty setting for the third term may not be a good idea for Strategy one as the third term absorbs too much of the image clutters that actually form the background, causing the background dictionary so constructed to lose representative power.
\\
In sum, for Strategy one, we set $\tau$ and $\lambda$ at $0.8$ and $0.133$ in the synthetic experiments, whereas at $3$ and $0.3$ in the real experiments. For Strategy two, the $\tau$ and $\lambda$ are set at $0.05$ and $0.02$ in the synthetic experiments, and $0.5$ and $0.2$ in the real experiments.

\vspace{-1.5mm}

\subsection{Synthetic Experiments}
\label{sec:synthetic_Experiments}
\vspace{-1.5mm}
%The experiments are done on a $101\times101$ zone from the acquired Cuprite scene.
%We incorporate in this zone, 7 target blocks (each of size $6\times 3$) with $\alpha\in[0.01, \, 1]$ (all have the same $\alpha$), placed in long convoy formation all formed by the same synthetic (perfect) target $\mathbf{t}$ consisting of a sulfate mineral type known as ''Jarosite''. We make sure by referring to Figure 5a in \cite{Swayze10245} that the small zone we consider does not already contain any Jarosite patches.
%The target $\mathbf{t}$ that we created actually consists of the mean of the first six Jarosite mineral samples taken from the United States Geological Survey (USGS - Reston) Spectral Library \cite{Clark93}. The target $\mathbf{t}$ replaces a fraction $\alpha\in[0.01, \, 1]$ from the background; specifically, the following values of $\alpha$ are considered: 0.01, 0.02, 0.05, 0.1, 0.3, 0.5, 0.8, and 1. As for $\mathbf{A}_t$, it is constructed from the six acquired Jarosite samples.

The experiments are done on a $101\times101$ zone (pixels in rows 389 to 489 and columns 379 to 479) from the acquired Cuprite scene.
We incorporate in this zone, 7 target blocks (each of size $6\times 3$) with $\alpha\in[0.01, \, 1]$ (all have the same $\alpha$), placed in long convoy formation all formed by the same synthetic (perfect) target $\mathbf{t}$ consisting of a sulfate mineral type known as ``Jarosite''. According to  Figure 5a in \cite{Swayze10245}, the small zone we consider here does not already contain any Jarosite patches. The target $\mathbf{t}$ that we created actually consists of the mean of the first six Jarosite mineral samples taken from the USGS Spectral Library \cite{Clark93}. The target $\mathbf{t}$ replaces a fraction $\alpha\in[0.01, \, 1]$ from the background; specifically, the following values of $\alpha$ are considered: 0.01, 0.02, 0.05, 0.1, 0.3, 0.5, 0.8, and 1. As for $\mathbf{A}_t$, it is constructed from the six acquired Jarosite samples.

\begin{figure}[!tbp]
~~~~~~~~~~~~$\mathbf{D}$ ~~~~~~~~~~= ~~~~~~~~~$\mathbf{L}$~~~~~~~~~~~  +~~~~~~$(\mathbf{A}_t\mathbf{C})^T$~~~~  +~~~~~~~~~~$\mathbf{N}$
\\
\minipage{0.12\textwidth}
  \includegraphics[width=\linewidth]{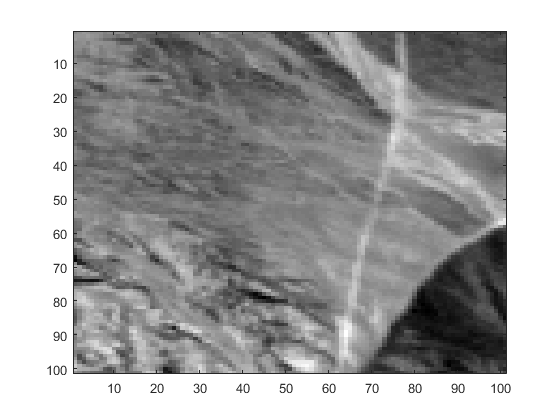}
\endminipage\hfill
\minipage{0.12\textwidth}
  \includegraphics[width=\linewidth]{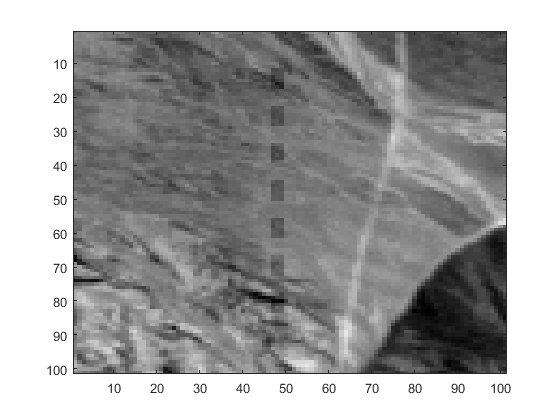}
\endminipage\hfill
\minipage{0.12\textwidth}
  \includegraphics[width=\linewidth]{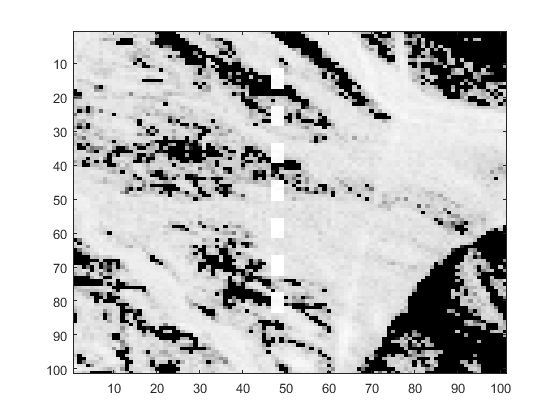}
\endminipage\hfill
\minipage{0.12\textwidth}
  \includegraphics[width=\linewidth]{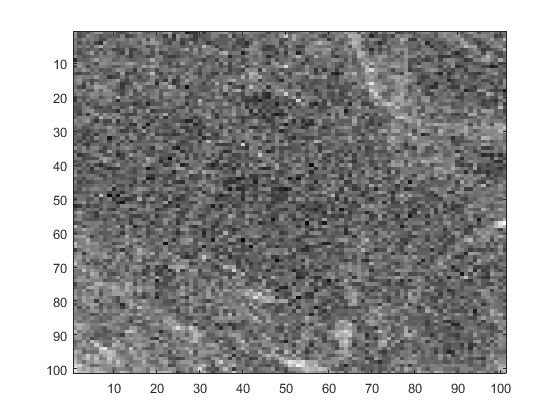}
\endminipage
\vspace{-4mm}
\caption{Visual separation of the 7 targets blocks for $\alpha=0.1$: We exhibit the mean power in dB over the 186 bands.}
%Column from left to right: the original HSI containing the 7 target blocks ($\alpha=0.1$), low rank background HSI $\mathbf{L}$, target HSI $(\mathbf{A}_t \mathbf{C})^T$, noise HSI.}
\label{fig:visual1}
\end{figure}

\begin{figure}[!tbp]
\vspace{-4mm}
\minipage{0.12\textwidth}
  \includegraphics[width=\linewidth]{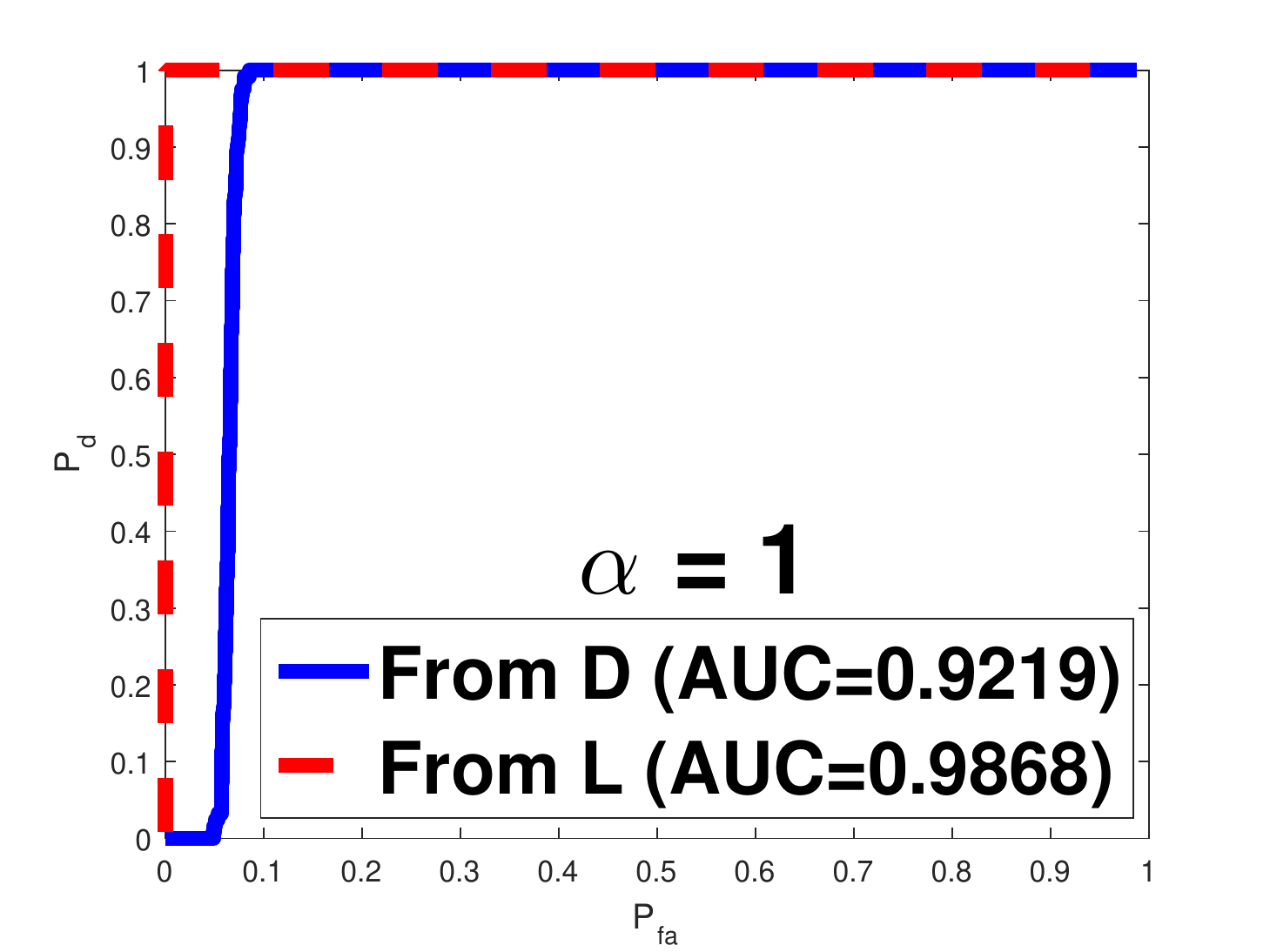}
\endminipage\hfill
\minipage{0.12\textwidth}
  \includegraphics[width=\linewidth]{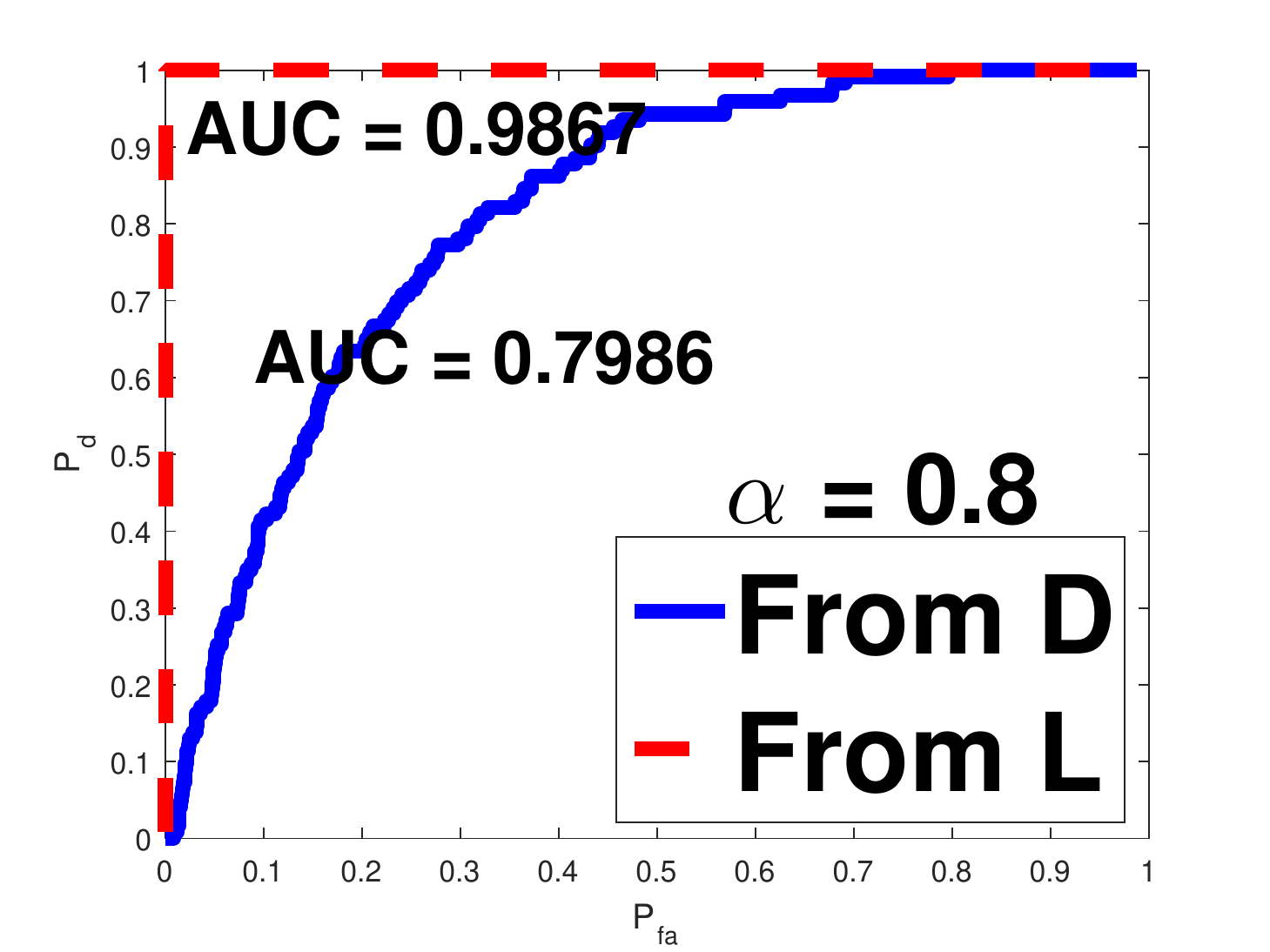}
\endminipage\hfill
\minipage{0.12\textwidth}
  \includegraphics[width=\linewidth]{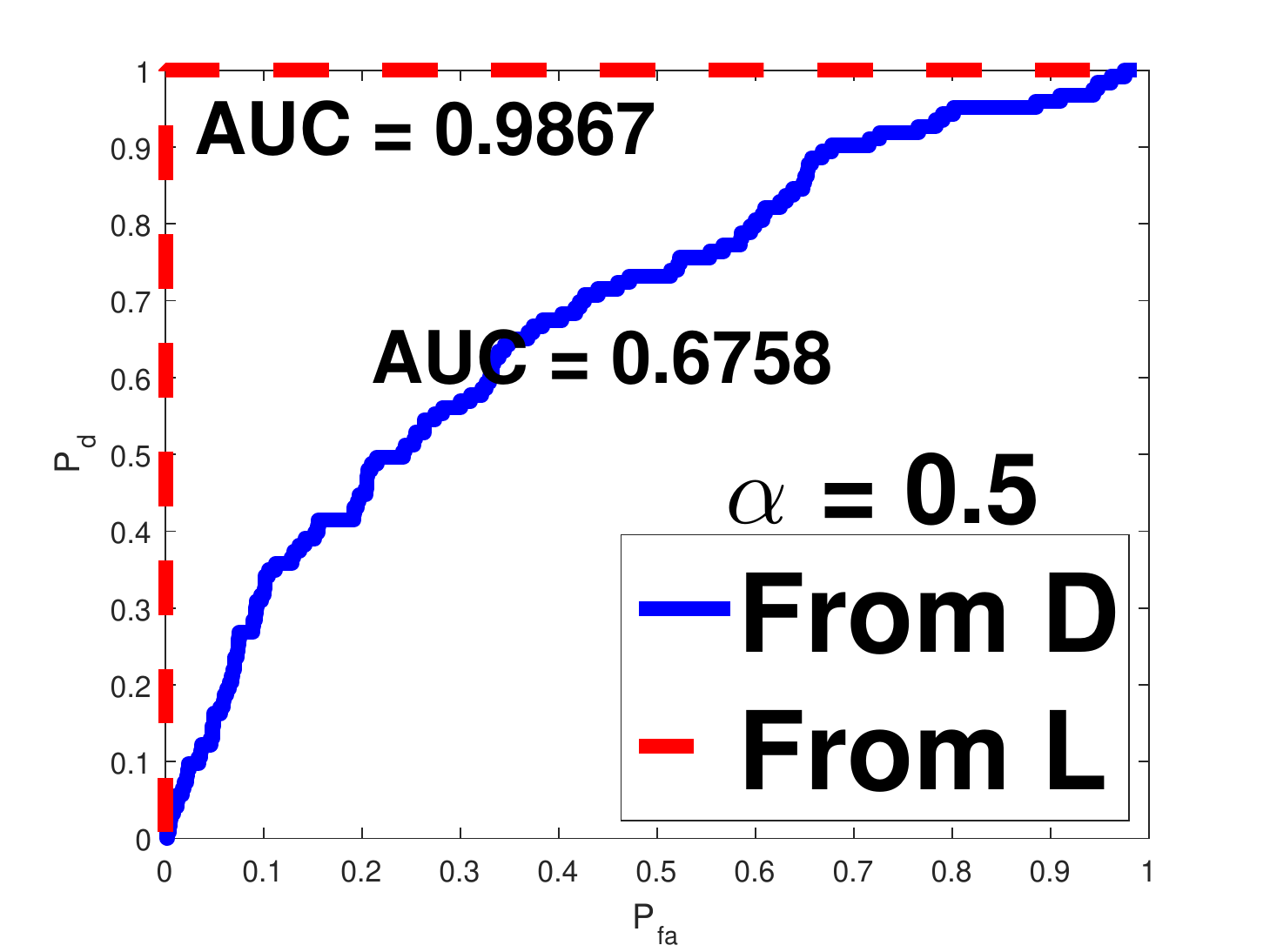}
\endminipage\hfill
\minipage{0.12\textwidth}
  \includegraphics[width=\linewidth]{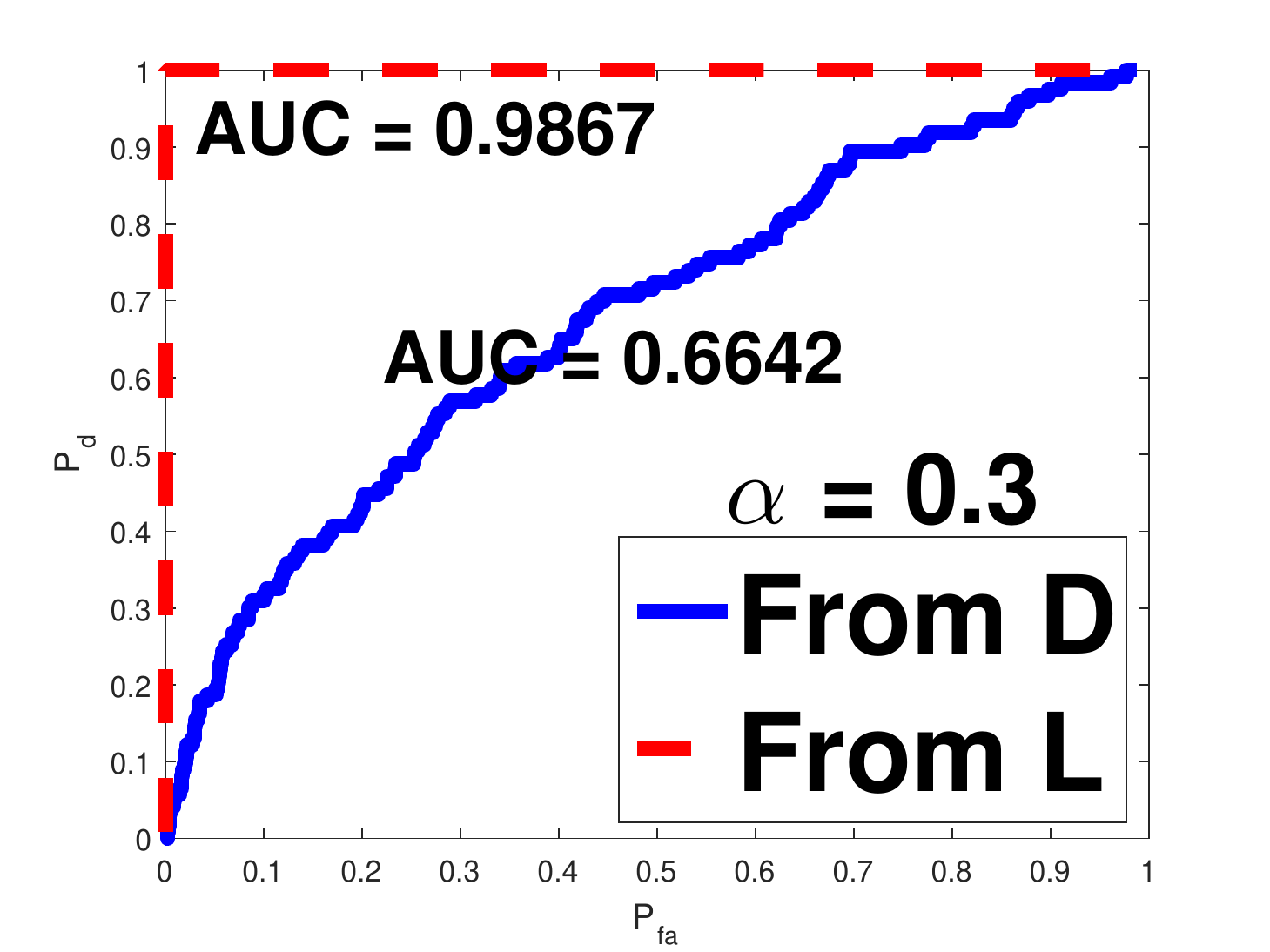}
\endminipage\hfill
\minipage{0.12\textwidth}
  \includegraphics[width=\linewidth]{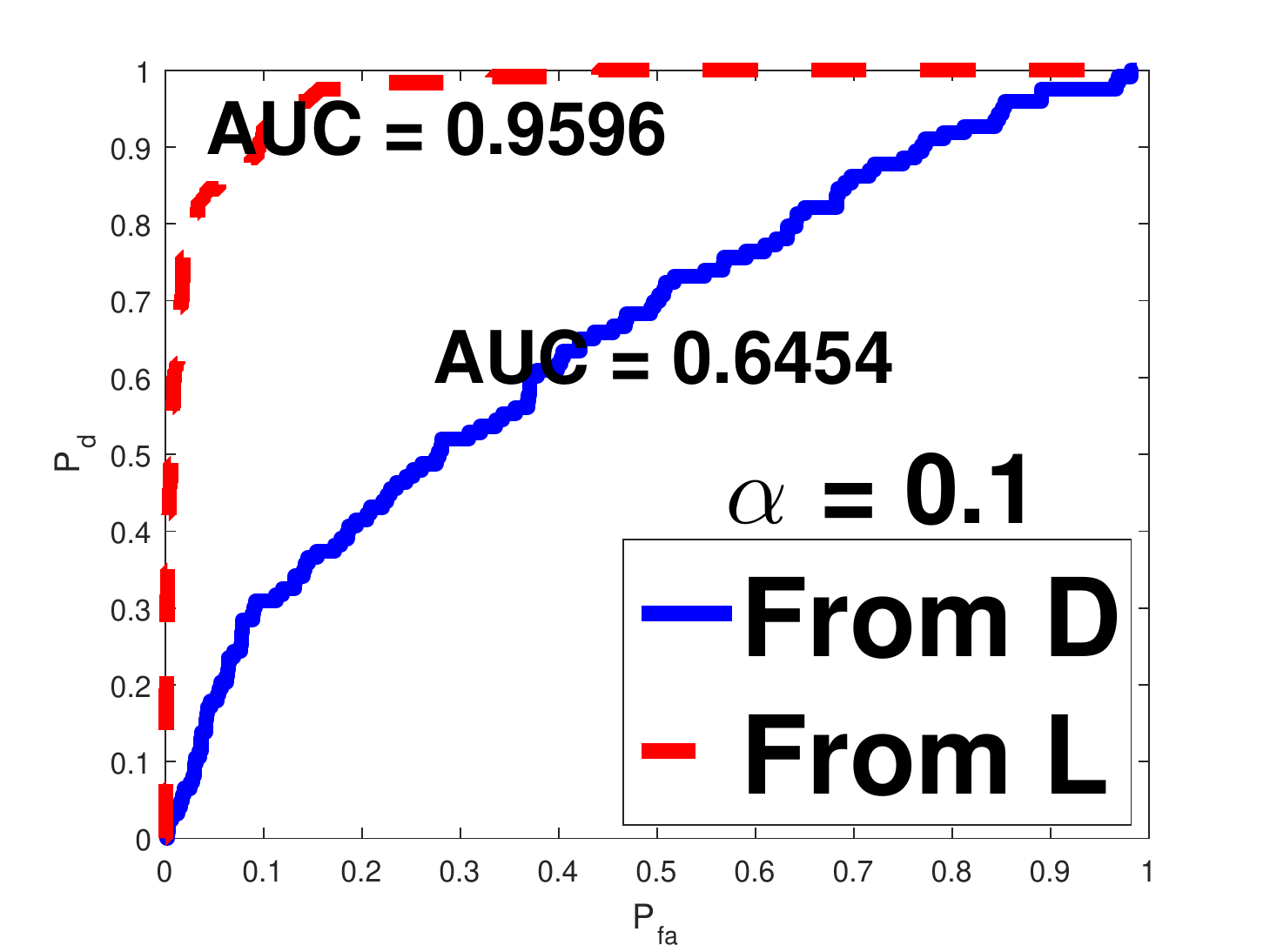}
\endminipage\hfill
\minipage{0.12\textwidth}
  \includegraphics[width=\linewidth]{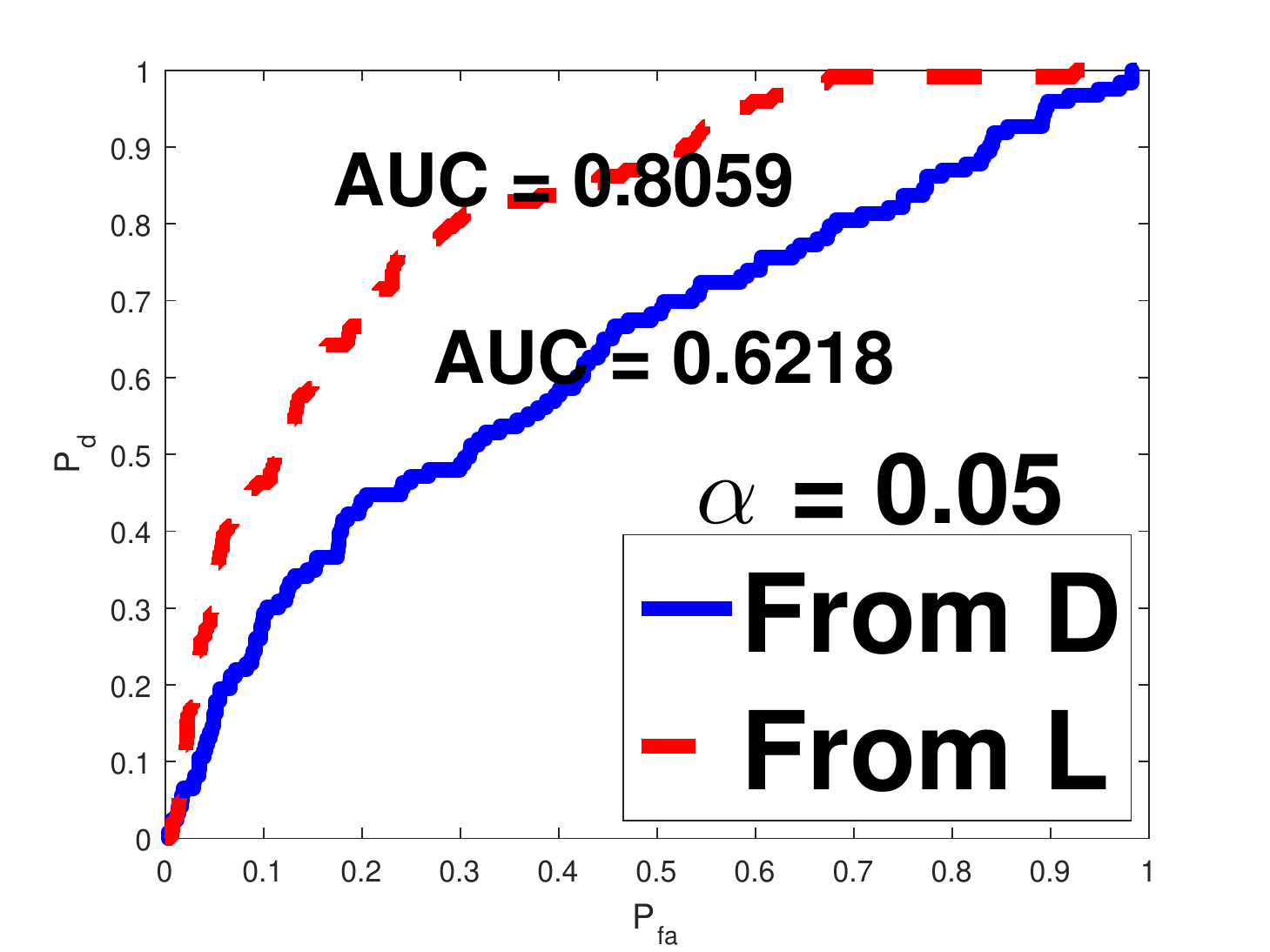}
\endminipage\hfill
\minipage{0.12\textwidth}
  \includegraphics[width=\linewidth]{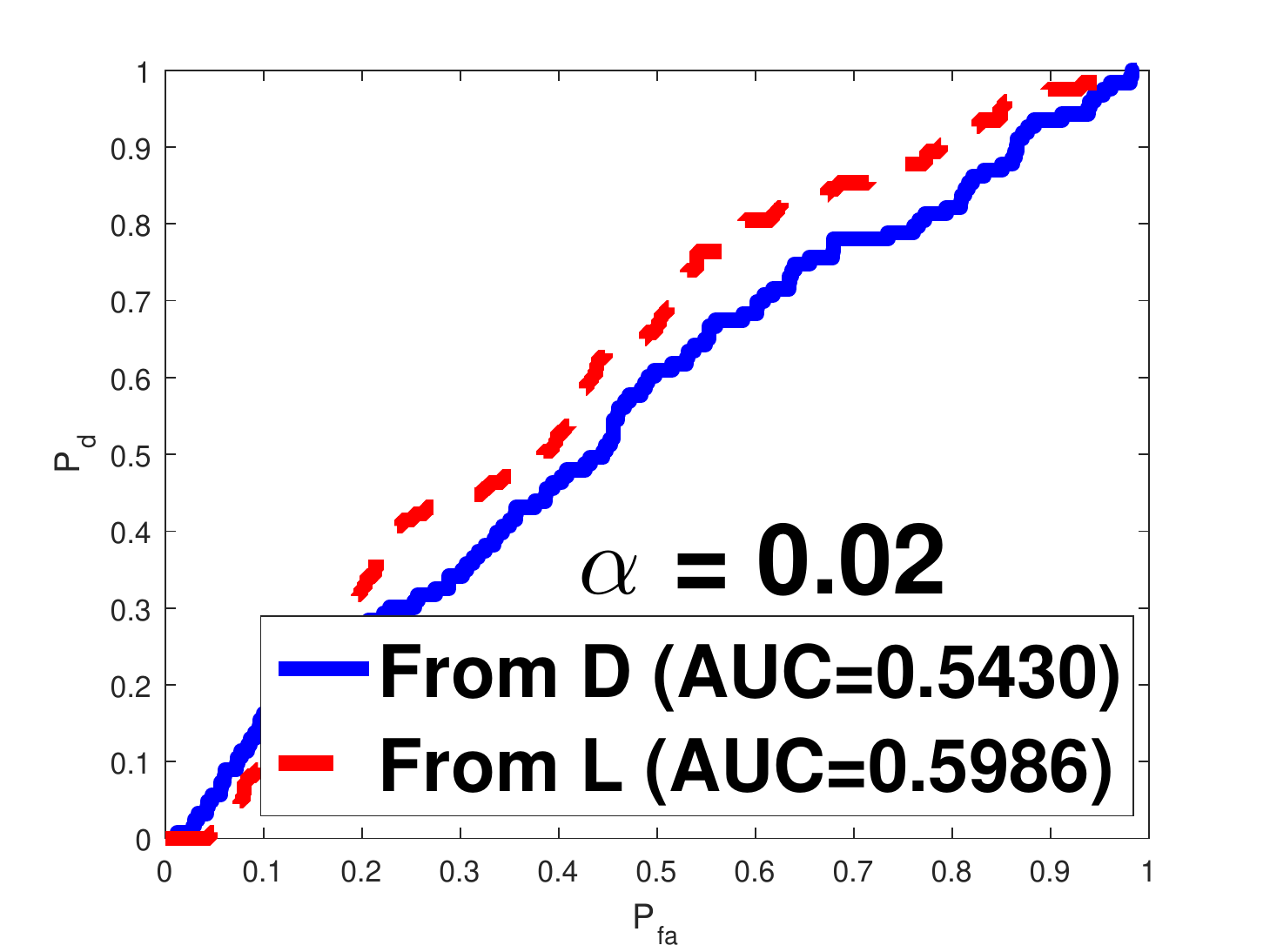}
\endminipage\hfill
\minipage{0.12\textwidth}
  \includegraphics[width=\linewidth]{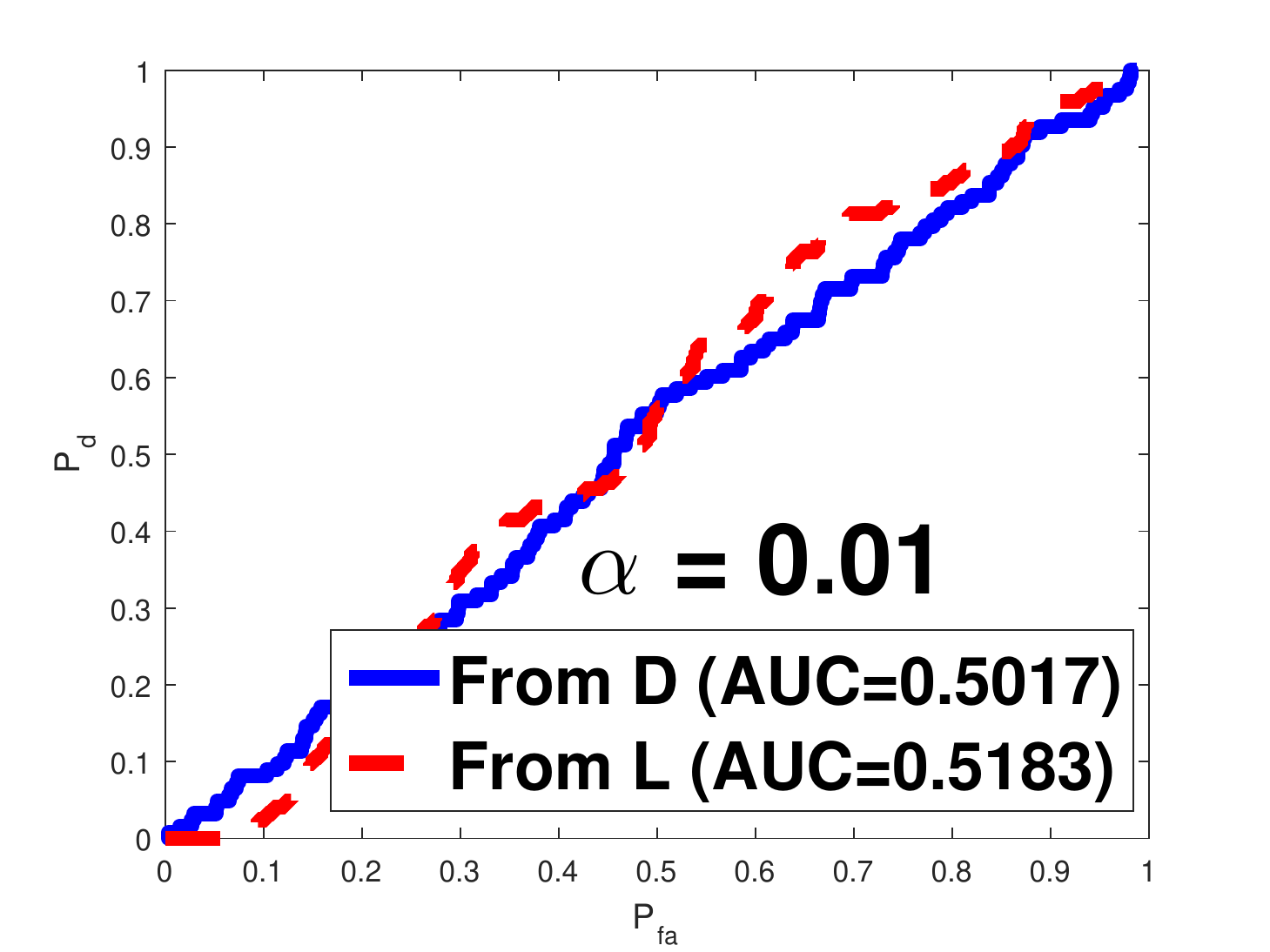}
\endminipage
\vspace{-4mm}
\caption{ROC curves (with their AUC values) for different $\alpha$ values of the SRBBH detector when $\mathbf{A}_b$ is constructed from $\mathbf{D}$ and $\mathbf{L}$.}
\label{fig:detec1}
\end{figure}

\vspace{-2.6mm}

\subsubsection{Using Strategy one for the detection}
\label{sec:Strategy_one_synthetic}
\vspace{-1.5mm}
We first provide in Figure \ref{fig:visual1} a visual evaluation of the separation of the 7 target blocks for low $\alpha=0.1$. We can observe that our problem \eqref{eq:convex_model} successfully discriminates these perceptually invisible targets from the background in $\mathbf{D}$ and separate them. The 7 darker blocks that appear in $\mathbf{L}$ correspond to the dimmer fraction of the background that remains after the targets have been removed at the corresponding spatial locations.
%In fact, we would be unrealistic if we expect that there would be a clean separation of the targets.

Having qualitatively inspect the separation, we now aim to quantitatively evaluate the target detection performances of the SRBBH detector \cite{Zhang15} when $\mathbf{A}_b$ is first constructed from $\mathbf{D}$ and then from $\mathbf{L}$ after applying problem \eqref{eq:convex_model}. We use a small concentric window of size $5\times 5$, and hence, $\mathbf{A}_b\in\mathbb{R}^{p\times24}$ (after excluding the center pixel). The detection performances are evaluated by the Receiver Operating Characteristics (ROC) curves and their corresponding Area Under Curve (AUC) values. 
%Note that by using this sliding concentric window, the image will be trimmed in function of the window size and hence the edges are not processed. Thus, to not lose a lot of pixels for testing, this is the main reason why we choose a small window size (that is, $5\times 5$).
Figure \ref{fig:detec1} depicts the quantitative detection results. Clearly, increasing $\alpha$ should render the target detection less challenging, and thus, better detection results are being expected. However, this fact can not always be the case for the SRBBH detector when $\mathbf{A}_b$ is constructed from $\mathbf{D}$: It is true that the increase in $\alpha$ helps to improve the detection, but at the same time leads to more target contamination in $\mathbf{A}_b$ which in turn suppresses the detection improvement that ought be had. That is why, the SRBBH detector (blue solid curves) does not reap full benefits from the increase in $\alpha$, and thus, presents poor detection results even for large $\alpha$ values. 
\\
By constructing $\mathbf{A}_b$ from $\mathbf{L}$, and due to the targets removal from the background after applying problem \eqref{eq:convex_model}, the SRBBH detector (dashed red curve) improves the detection especially for $\alpha\geq 0.1$. The detection performances start to deteriorate progressively for very small $\alpha$ values and degenerate to the SRBBH level (blue solid curve) for $\alpha\leq0.02$. 
To sum up, the obtained target detection results corroborate our claim that we can handle targets with low fill-fraction (e.g. in camouflage) and in convoy formation.

\vspace{-2.6mm}

\subsubsection{Using Strategy two for the detection}
\begin{figure}[!tbp]
\minipage{0.12\textwidth}
  \includegraphics[width=\linewidth]{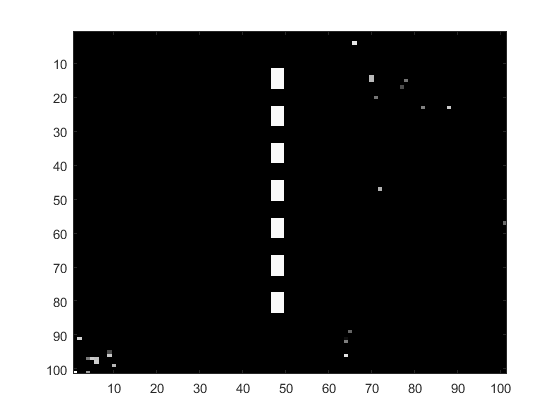}
\endminipage\hfill
\minipage{0.12\textwidth}
  \includegraphics[width=\linewidth]{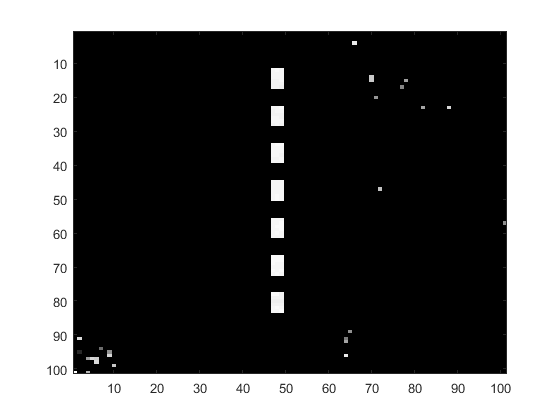}
\endminipage\hfill
\minipage{0.12\textwidth}
  \includegraphics[width=\linewidth]{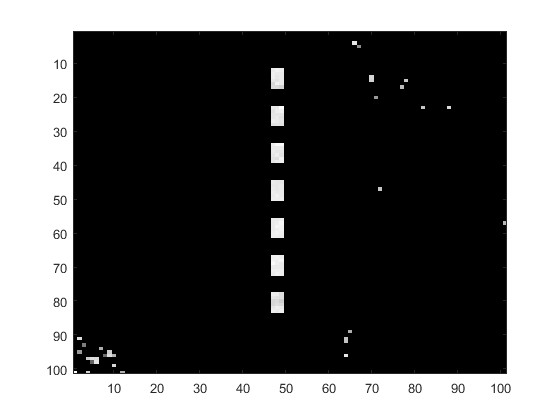}
\endminipage\hfill
\minipage{0.12\textwidth}
  \includegraphics[width=\linewidth]{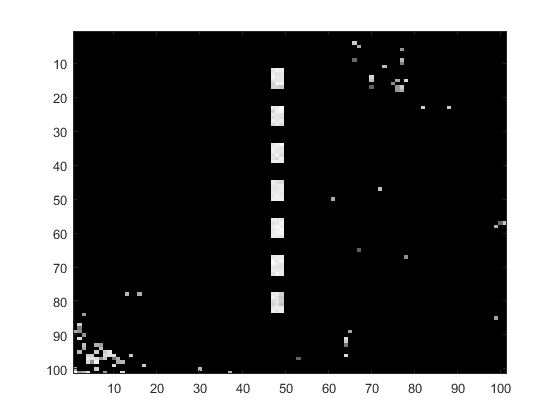}
\endminipage\hfill
\minipage{0.12\textwidth}
  \includegraphics[width=\linewidth]{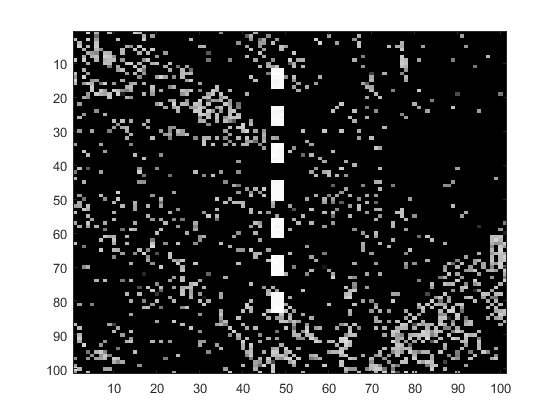}
\endminipage\hfill
\minipage{0.12\textwidth}
  \includegraphics[width=\linewidth]{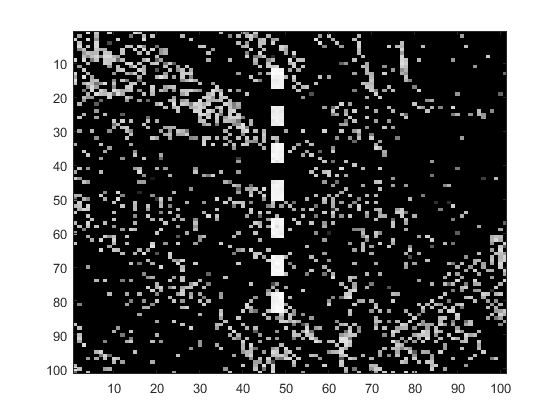}
\endminipage\hfill
\minipage{0.12\textwidth}
  \includegraphics[width=\linewidth]{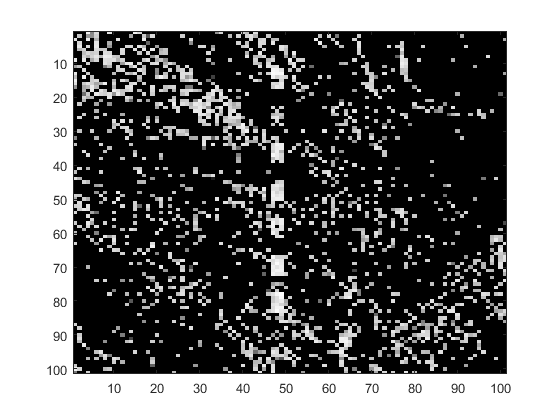}
\endminipage\hfill
\minipage{0.12\textwidth}
  \includegraphics[width=\linewidth]{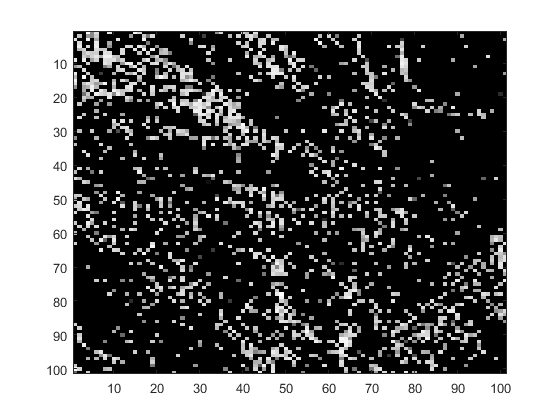}
\endminipage
\vspace{-4mm}
\caption{Visual detections (mean power in dB over the 186 bands) of $(\mathbf{A}_t\mathbf{C})^T$ for the 7 target blocks for different $\alpha$ values. From the top left to the bottom right for $\alpha$: 1, 0.8, 0.5, 0.3, 0.1, 0.05, 0.02, 0.01.}
\label{fig:visual3}
\end{figure}

\vspace{-1.5mm}
Figure \ref{fig:visual3} depicts a 2-D visual detection results of $(\mathbf{A}_t\mathbf{C})^T$ for different $\alpha$ values. Obviously, Strategy two detects all the targets with little false alarms until $\alpha \leq0.1$ when a lot of false alarms appear.

%decreasing $\alpha$ the false alarms increases at each decrease step value of $\alpha$. This is to be very expected since 
%According Strategy two in Subsection \ref{sec:what_after}, here we only interest to automatically locate the targets of interests that reside in the target subspace specified by $\mathbf{A}_t$ in the target HSI.
%Recall that here, we need that $\lambda$ to be high enough. Here we follow the same procedure we defined in \ref{sec:Strategy_one_synthetic}

\vspace{-1.5mm}
\subsection{Real Experiments}
\label{sec:global_real_Experiments}
\vspace{-1.5mm}
The experiments are based on a region of size 250 $\times$ 291 pixels (see Figure {\bf1(b)}) taken from the acquired Cuprite HSI. We consider this zone specifically to detect the Tectosilicate mineral type target pixels known as Buddingtonite. There are three Buddingtonite samples available in the online ASTER spectral library \cite{Baldridge09}, and our target dictionary $\mathbf{A}_t$  is formed by these samples.

\begin{figure}[!tbp]
\vspace{-4mm}
\minipage{0.161\textwidth}
  \includegraphics[width=\linewidth]{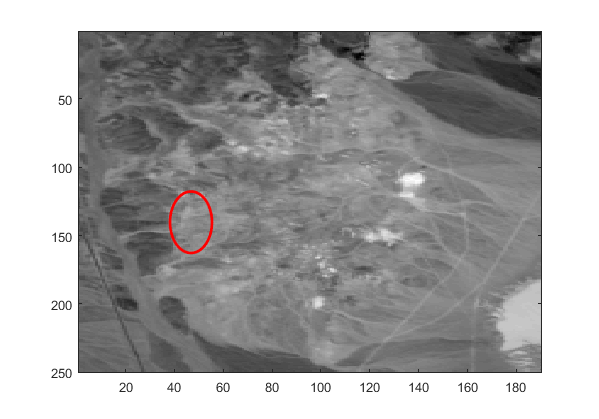}
\endminipage\hfill
\minipage{0.161\textwidth}
  \includegraphics[width=\linewidth]{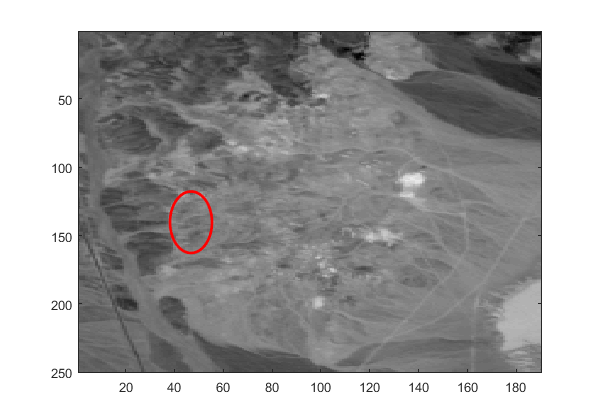}
\endminipage\hfill
\minipage{0.150\textwidth}
  \includegraphics[width=\linewidth]{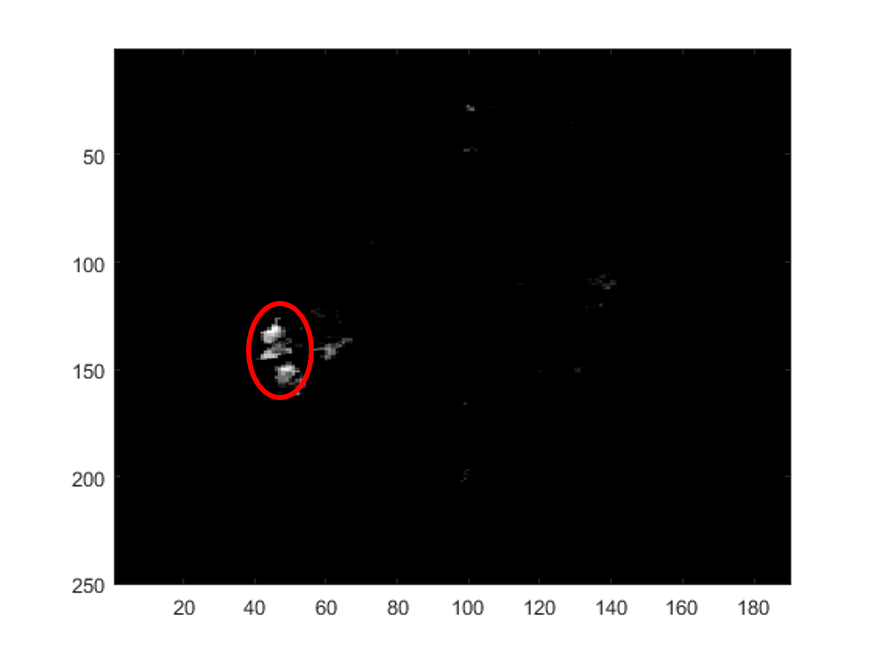}
\endminipage
\vspace{-4mm}
\caption{Visual separation (mean power in dB over the 186 bands) of the Buddingtonite targets. Columns from left to right: original HSI, low rank background HSI $\mathbf{L}$, $(\mathbf{A}_t\mathbf{C})^T$ after some thresholding.}\label{fig:separation_Buddingtonite}\label{fig:Buddingtonite_separation_ourMethod}
\end{figure}

\begin{figure}[!htb]
\minipage{0.16\textwidth}
  \includegraphics[width=\linewidth]{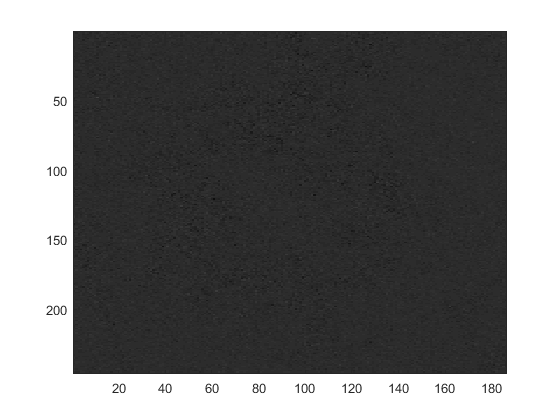}
\endminipage\hfill
\minipage{0.16\textwidth}
  \includegraphics[width=\linewidth]{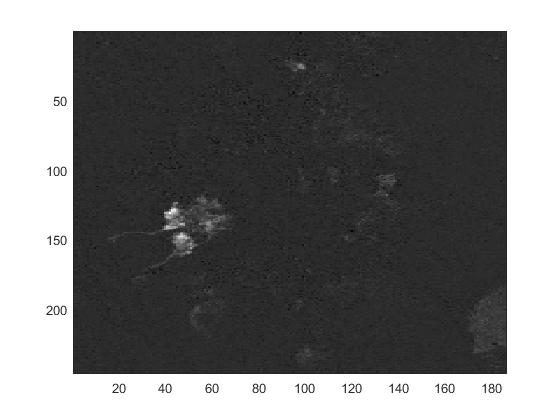}
\endminipage\hfill
\minipage{0.16\textwidth}
  \includegraphics[width=\linewidth]{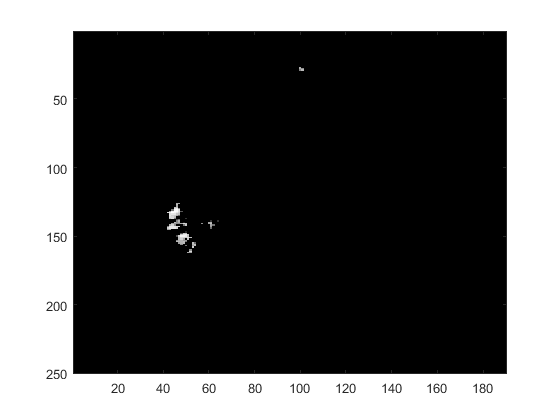}
\endminipage
\vspace{-4mm}
  \caption{Visual detection of the Buddingtonite target pixels (2-D display). Columns from left to right: SRBBH detector when $\mathbf{A}_b$ is constructed from $\mathbf{D}$, SRBBH detector when $\mathbf{A}_b$ is constructed from $\mathbf{L}$, Detection in $(\mathbf{A}_t\mathbf{C})^T$ for Strategy two (mean power in dB).}\label{fig:final_results}
\end{figure}

{\em Using Strategy one}: As a consequence of the decomposition depicted in Figure \ref{fig:Buddingtonite_separation_ourMethod}, the overlap problem illustrated in Figure \ref{fig:SPCP_example} is now much relieved, as can also be observed from Figure \ref{fig:final_results}. The first and second column in Figure \ref{fig:final_results} evaluate visually the SRBBH detection results when $\mathbf{A}_b$ is constructed from $\mathbf{D}$ and $\mathbf{L}$, respectively, using a concentric window of size $5\times 5$. We can obviously observe the effectiveness of problem \eqref{eq:convex_model} in improving the target detection.
%Now, by visually evaluating the SRBBH detector (using a concentric window of size $5\times 5$) when $\mathbf{A}_b$ is first constructed from $\mathbf{D}$ and then from $\mathbf{L}$, the first two columns in \ref{fig:final_results} demonstrate the effectiveness of problem \eqref{eq:convex_model} in potentially improving the detection for the SRBBH.

{\em Using Strategy two}: The third column in Figure \ref{fig:final_results} depicts the visual detection of the Buddingtonite targets in $(\mathbf{A}_t\mathbf{C})^T$. The Buddingtonite targets are detected with very little false alarms.
 %\footnote{Both th HSI and the Buddingtonite target samples (and hence, all the atoms in $\mathbf{A}_t$) are normalized to values between 0 and 1.}

%\subsubsection{Using \underline{Strategy two} for the detection}
%By progressively raising the abolute level of $\tau$ to $\lambda$ under a ratio of $1:4$, we noticed that until a crertain limit, the false alarms (specially the image clutters that correspond to the small high contrast regions present) disappear until the Buddingtonite

\vspace{-4mm}

\vspace{-7mm}
\section{Conclusion and Future work}
\vspace{-1.5mm}
A method based on a modification of RPCA is proposed for separating the targets from the background in hyperspectral imagery. Two strategies are briefly outlined to realize the target detection. 
In the future, we will use other proxies than the $l_{2,1}$ norm (closer to $l_{2,0}$) to alleviate the $l_{2,1}$ artifact and the manual selection problem of $\tau$, $\lambda$.

\vspace{-3mm}
\section{Acknowledgment}
\vspace{-1.5mm}
The authors would like to thank Dr.~Gregg A. Swayze from the USGS Spectroscopy Lab for his time in providing us helpful remarks and suggestions specially about the Cuprite data. 

% Below is an example of how to insert images. Delete the ``\vspace'' line,
% uncomment the preceding line ``\centerline...'' and replace ``imageX.ps''
% with a suitable PostScript file name.
% -------------------------------------------------------------------------
%\begin{figure}[htb]

%\begin{minipage}[b]{1.0\linewidth}
 % \centering

%The authors would like to thank Dr. Gregg Swayze for his 
%  \centerline{\includegraphics[width=8.5cm]{image1}}
%  \vspace{2.0cm}
  %\centerline{(a) Result 1}\medskip
%\end{minipage}
%
%\begin{minipage}[b]{.48\linewidth}
 % \centering
  %\centerline{\includegraphics[width=4.0cm]{image3}}
%  \vspace{1.5cm}
%  \centerline{(b) Results 3}\medskip
%\end{minipage}
%\hfill
%\begin{minipage}[b]{0.48\linewidth}
  %\centering
  %\centerline{\includegraphics[width=4.0cm]{image4}}
%  \vspace{1.5cm}
  %\centerline{(c) Result 4}\medskip
%\end{minipage}
%
%\caption{Example of placing a figure with experimental results.}
%\label{fig:res}
%
%\end{figure}

%\vfill
%\pagebreak

\vfill\pagebreak
%\bibliographystyle{IEEEbib}
%\bibliography{strings,refs}
\footnotesize
\bibliographystyle{IEEEtran}
\bibliography{biblio_these2}

\end{document}